\documentclass[aps,prl,superscriptaddress,reprint,floatfix,showpacs]{revtex4-1}
\usepackage{amsmath,bm} 
\usepackage{graphicx}
\usepackage{color}

\begin{document}

\title{Breakdown of the Peierls substitution for the Haldane model with ultracold atoms}

\author{Julen Iba\~nez-Azpiroz}
\affiliation{\mbox{Peter Gr\"unberg Institute and Institute for Advanced Simulation, Forschungszentrum J\"ulich \& JARA, D-52425 J\"ulich, Germany}}

\author{Asier Eiguren}
\affiliation{Depto. de F\'isica de la Materia Condensada, Universidad del Pais Vasco, UPV/EHU, 48080 Bilbao, Spain}
\affiliation{Donostia International Physics Center (DIPC), 20018 Donostia, Spain}

\author{Aitor Bergara}
\affiliation{Depto. de F\'isica de la Materia Condensada, Universidad del Pais Vasco, UPV/EHU, 48080 Bilbao, Spain}
\affiliation{Donostia International Physics Center (DIPC), 20018 Donostia, Spain}
\affiliation{Centro de F\'{i}sica de Materiales CFM, Centro Mixto CSIC-UPV/EHU, 20018 Donostia, Spain}

\author{Giulio Pettini}
\affiliation{Dipartimento di Fisica e Astronomia, Universit\`a di Firenze,
and INFN, 50019 Sesto Fiorentino, Italy}

\author{Michele Modugno}
\affiliation{\mbox{Depto. de F\'isica Te\'orica e Hist. de la Ciencia, Universidad del Pais Vasco UPV/EHU, 48080 Bilbao, Spain}}
\affiliation{IKERBASQUE, Basque Foundation for Science, 48011 Bilbao, Spain}

\begin{abstract}
We present two independent calculations of the tight-binding parameters for a specific realization of the Haldane model with ultracold atoms. The tunneling coefficients up to next-to-nearest neighbors are computed \textit{ab-initio} by using the maximally localized Wannier functions, and compared to analytical expressions written in terms of gauge invariant, measurable properties of the spectrum. The two approaches present a remarkable agreement and  evidence the breakdown of the Peierls substitution: \textit{(i)} the phase acquired by the next-to-nearest tunneling amplitude $t_{1}$ presents quantitative and qualitative differences with respect to that obtained by the integral of the vector field $\bm{A}$, as considered in the Peierls substitution, even in the regime of low amplitudes of $\bm{A}$; \textit{(ii)} for larger values, also $|t_{1}|$ and the nearest-neighbor tunneling $t_{0}$ have a marked dependence on $\bm{A}$. The origin of this behavior and its implications are discussed.
\end{abstract}

\date{\today}

\pacs{67.85.-d, 73.43.-f}
\maketitle

The so-called \textit{Peierls substitution}, named after the original work by R. Peierls \cite{peierls1933}, is a widely employed approximation for describing tight-binding electrons in the presence of a slowly varying external vector field. It is usually encountered in either of these two forms, as a modification of the semiclassical dispersion, $E(\bm{k})\to E(-i\hbar\bm{\nabla} -(e/c)\bm{A})$ \cite{luttinger1951}, or as a phase factor acquired by the tunneling amplitudes of  the corresponding tight-binding Hamiltonian, $t_{ij}\to t_{ij}\exp\{ie\int_{i}^{j}\bm{A}d\bm{r}\}$ \cite{bernevig2013}. The latter expression must be evaluated on the straight path connecting sites $i$ and $j$, as demonstrated under the hypothesis of a same-site, same-orbital interaction with the vector field by Boykin \textit{et al.} \cite{boykin2001}. 

Despite its popularity, the Peierls substitution is a rather uncontrolled approximation, as already pointed out in Refs. \cite{alexandrov1991,*alexandrov1991a,kohn1959}. For example, we notice that the integral of the vector field appearing in the Peierls phase factor has been conventionally taken along a straight path (see e.g. \cite{haldane1988,graf1995}) long before its formal demonstration \cite{boykin2001}, just for convenience (in principle, in two and three dimensions there is an ambiguity as the path is not univocally defined \cite{bernevig2013}). In addition, in the literature the Peierls  substitution is often applied as a ``magic formula'', with little care about its regime of validity.

The Peierls substitution plays a fundamental role in the Haldane model \cite{haldane1988}, a celebrated two-dimensional periodic tight-binding model, characterized by a quantum Hall effect caused by the breaking of time-reversal symmetry with zero magnetic flux through the unit cell \cite{haldane1988}. The model is characterized by exotic quantum phases, with different Chern numbers, depending on the value of the phase $\varphi$ of the next-to-nearest tunneling amplitude $t_{1}$, that is usually computed by the integral of the vector field $\bm{A}$ cited above.
Recently, in the literature there have been proposals for engineering the Haldane model with ultracold atoms in optical lattices by means of artificial gauge fields \cite{shao2008,anisimovas2014}, and to study the associated topological quantum states in the presence of sharp boundaries \cite{stanescu2009,*stanescu2010}. In fact, these systems represent a very interesting platform for simulating solid state physics \cite{lewenstein2012}. Again, these studies make use of approximate methods to deal with the tunneling amplitudes, by exploiting the Peierls substitution \textit{tout court}  \cite{shao2008} (see also \cite{goldman2009,williams2010,struck2012,jimenez-garcia2012,hauke2012,mazzucchi2013,aidelsburger2013,struck2013,polak2013}), or by using approximate atomic orbitals \cite{stanescu2009}.

In this Letter we present two independent calculations of the tight-binding parameters for the Haldane model  discussed in Refs. \cite{shao2008,stanescu2009}. In particular, we show that, within the next-to-nearest neighbors approximation, the tunneling coefficients can be directly written in terms of gauge invariant, measurable properties of the spectrum (namely the gap at the Dirac point and the bandwidths), \textit{or} computed  \textit{ab-initio} by using the maximally localized Wannier functions (MLWFs) \cite{marzari1997,*marzari2012,ibanez-azpiroz2013,ibanez-azpiroz2013a}. Notably, the two approaches present a remarkable agreement, evidencing the breakdown of the Peierls substitution. As a matter of fact, the phase acquired by the next-to-nearest tunneling amplitude $t_{1}$ is quantitatively different from that predicted by the integral of the vector field $\bm{A}$, and presents a pronounced dependence on the intensity of the underlying scalar potential. Moreover, both the amplitude of $t_{1}$ and of the nearest-neighbor tunneling $t_{0}$ turn out to be dependent on the intensity of $\bm{A}$.

\begin{figure}
\centerline{\includegraphics[width=0.9\columnwidth]{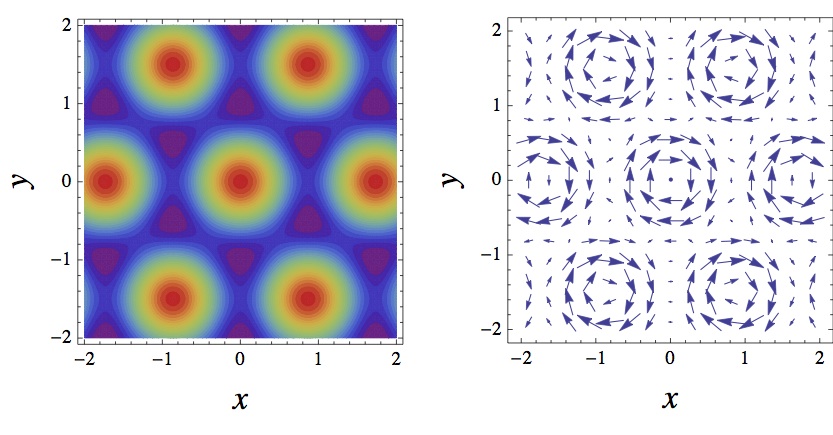}}
\centerline{\includegraphics[width=0.9\columnwidth]{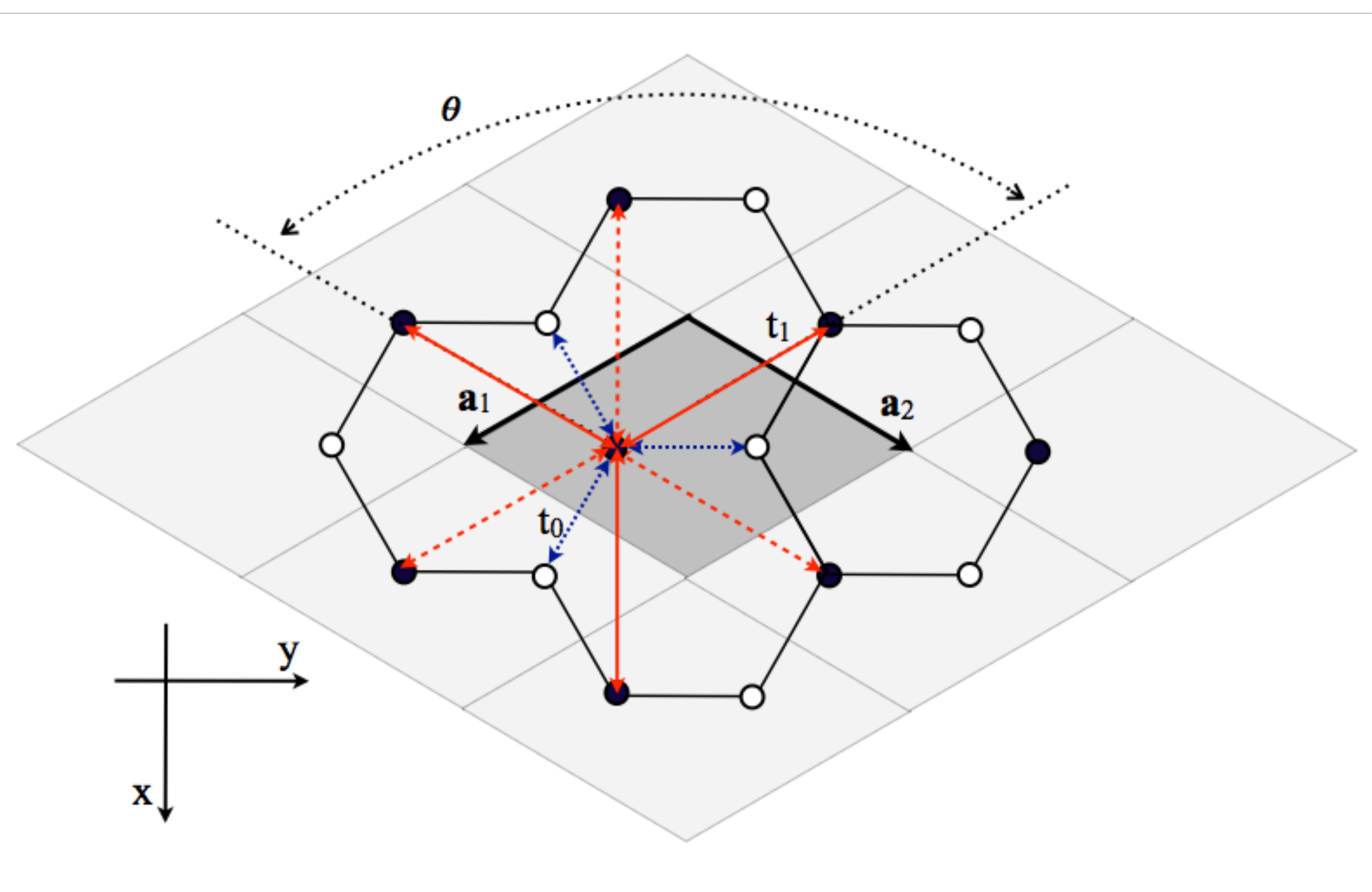}}
\caption{(Color online) Top: Structure of the scalar (left) and vector (right) potentials. In the left panel, hot and cold colors correspond to maxima and minima of the potential, respectively.
Bottom: Bravais lattice associated to the honeycomb potential in Eq. (\ref{eq:pot}). Black and white circles refers to minima of type $A$ and $B$, respectively. The elementary cell is highlighted in gray. The various tunneling coefficients are indicated for the site of type $A$ in the central cell. 
The system is invariant under discrete translation generated by the Bravais vectors $\bm{a}_{1/2}$ and under rotations by $\theta=2\pi/3$ radians around any vertex of the lattice.
The former implies that next-to-nearest tunneling amplitudes $t_{1}$ along the same direction are conjugate pairs (solid and dashed lines); from the latter follows the equivalence of the hopping amplitudes separated by 
$2\pi/3$ radians. When sites $A$ and $B$ are degenerate, the system is also invariant 
under rotations by $\pi$ radians around the center of any elementary cell; this implies that $t_{0}$ is real.}
\label{fig:honeycomb}
\end{figure}

Let us start from the following single-particle, minimal-coupling Hamiltonian in two-dimensions
\begin{equation}
\hat{H_{0}}=\frac{1}{2m}\left[\hat{\bm{p}}-\bm{A}(\bm{r})\right]^{2} + V_{L}(\bm{r})
\label{eq:hamiltonian}
\end{equation}
with $\bm{r}=(x,y)$, $\bm{p}=-i\hbar\nabla$, and $V_{L}$ being the following honeycomb potential \cite{stanescu2009,lee2009,ibanez-azpiroz2013} 
\begin{equation}
V_{L}(\bm{r})=sE_{R}\!\left[2\cos\left((\bm{b}_{1}-\bm{b}_{2})\!\cdot\!\bm{r}\right) + 2\!\sum_{i=1}^{2}\cos\left(\bm{b}_{i}\!\cdot\!\bm{r}\right)\right]
\label{eq:pot}
\end{equation}
where the vectors $\bm{b}_{1/2}=({\sqrt{3}}/{2}k_{L}) (\bm{e}_{x}\mp\sqrt{3}{\bm{e}}_{y})$ generate the reciprocal lattice, $k_{L}$ is the laser wavelength and $s$ the amplitude of the potential 
in units of the recoil energy $E_{R}=\hbar^{2}k_{L} ^{2}/2m$ \footnote{The potential can be exactly mapped into that used in our previous work \cite{ibanez-azpiroz2013}, by a counterclockwise rotation by $90^{\circ}$ of the axes ($\bm{e}_{x}\to\bm{e}_{y}$, $\bm{e}_{y}\to-\bm{e}_{x}$). The same potential is also equivalent to that used by Stanescu \textit{et al.} \cite{stanescu2009} (by posing $V_{0}=4s$) and to that by Shao \textit{et al.} \cite{shao2008} (except for an irrelevant shift of the coordinates).}. Notice that, though this specific realization is characterized by degenerate potential wells, an imbalance can be easily produced by introducing a suitable phase \cite{lee2009,shao2008}.
The corresponding Bravais lattice, 
${\cal B}=\{j_{1}{\bf{a}}_{1}+j_{2}{\bf{a}}_{2}\Big| j_{1},j_{2}=0,\pm 1,\pm 2 \dots\}$,
with lattice constant $a$ (such that $k_{L}=4\pi/(3\sqrt{3}a)$ \cite{lee2009}),
is generated by the two basis vectors 
$\bm{a}_{1/2} =({2\pi}/{3k_{L}}) (\bm{e}_{x},\mp\sqrt{3}\bm{e}_{y})$, obeying $\bm{a}_i\!\cdot\!\bm{b}_j=2\pi\delta_{ij}$.
As for the vector potential, we consider the same expression discussed in Refs. \cite{shao2008,stanescu2009} (corresponding to the Coulomb gauge, $\nabla\!\cdot\!\bm{A}(\bm{r})=0$) 
\begin{align}
&\bm{A}(\bm{r})=\alpha\hbar k_{L}\left[\left(\sin((\bm{b}_{2}-\bm{b}_{1})\!\cdot\!\bm{r}) 
+ \frac12\sin(\bm{b}_{2}\!\cdot\!\bm{r})
\right.\right.
\\
&\qquad
\left.\left.-\frac12\sin(\bm{b}_{1}\!\cdot\!\bm{r})\right)\bm{e}_{x}
-\frac{\sqrt{3}}{2}\left(\sin\left(\bm{b}_{1}\!\cdot\!\bm{r}\right)
+\sin\left(\bm{b}_{2}\!\cdot\!\bm{r}\right)\right)\bm{e}_{y}\right]
\nonumber
\end{align}
that has the same symmetry of the underlying honeycomb potential (see Fig. \ref{fig:honeycomb}).
The parameter $\alpha$ represents the amplitude of the vector potential in units of $\hbar k_{L}$.

The tight-binding model is constructed from the many-body Hamiltonian $\hat{\cal{H}}_{0}=\int d\bm{r}~{\hat{\psi}}^\dagger(\bm{r})\hat{H}_{0}{\hat{\psi}}(\bm{r})$, 
by expanding the field operator on a basis of localized functions, 
$\hat{\psi}(\bm{r})\equiv \sum_{\bm{j}\nu}{\hat{a}}_{\bm{j}\nu}w_{\bm{j}\nu}(\bm{r})$,
with the usual commutation rules $[{\hat{a}}_{\bm{j}\nu},{\hat{a}}^{\dagger}_{\bm{j'}\nu'}]=\delta_{\bm{jj'}}\delta_{\nu\nu'}$. Then,  by restricting the analysis to the two lowest bands, $\hat{\cal{H}}_{0}$ can be written as \cite{ibanez-azpiroz2013,ibanez-azpiroz2013a}
\begin{equation}
\hat{\cal{H}}_0 = \sum_{\nu\nu'=A,B}\sum_{\bm{j,j'}}{\hat{a}}^{\dagger}_{\bm{j}\nu}{\hat{a}}_{\bm{j}'\nu'}
\langle w_{\bm{j}\nu}|{\hat{H}}_0|w_{\bm{j'}\nu'}\rangle,
\label{singparth0}
\end{equation}
where the matrix elements $\langle w_{\bm{j}\nu}|{\hat{H}}_0|w_{\bm{j'}\nu'}\rangle$ correspond to tunneling amplitudes between different lattice sites
(except for the special case $\bm{j'}=\bm{j}$, $\nu=\nu'$, representing the onsite energies). These matrix elements depend only on $\bm{j'-j}$ due to the translational invariance of the lattice. The spectrum of $\hat{\cal{H}}_0$ can be obtained by considering the following transformation from coordinate to momentum space,
$\hat{b}_{\nu{\bm{k}}}=({1}/{\sqrt{S_{\cal  B}}})
\sum_{\bm{j}} ~e^{-i{\bm{k}}\cdot{\bm{R}}_{\bm{j}}}\hat{a}_{\bm{j}{\nu}}$, yielding \begin{equation}
\hat{\cal{H}}_{0}^{tb}=\sum_{\nu\nu'=A,B}\int_{S_{\cal B}} d\bm{k}~h_{\nu\nu'}(\bm{k})
\hat{b}_{\nu\bm{k}}^{\dagger}\hat{b}_{\nu'\bm{k}},
\end{equation}
with $h_{\nu\nu'}({\bm{k}})=\sum_{\bm{j}}e^{i{\bm{k}}\cdot{\bm{R}}_{\bm{j}}}\langle {w}_{\bm{0}\nu}|\hat{H}_{0}|w_{\bm{j}\nu'}\rangle$, and $S_{\cal  B}$ indicating the first Brillouin zone \cite{ibanez-azpiroz2013,ibanez-azpiroz2013a}.
By truncating the above expression to next-to-nearest neighbors as usual \cite{haldane1988,shao2008}, we define
\begin{equation}
h_{\nu\nu'}({\bm{k}})\equiv \left[h^{(0)}_{\nu\nu}(\bm{k})+h^{(2)}_{\nu\nu}(\bm{k})\right]\delta_{\nu\nu'}+h^{(1)}_{\nu\nu'}(\bm{k}).
\end{equation}
The first term corresponds to the onsite energies,
\begin{equation}
h^{(0)}_{\nu\nu}(\bm{k})=\langle {w}_{\bm{0}\nu}|\hat{H}_{0}|w_{\bm{0}\nu}\rangle\equiv E_{\nu}.
\end{equation}
The second term has only off-diagonal elements, corresponding to the hopping toward the three nearest-neighbor sites (see Fig. \ref{fig:honeycomb}). Thanks to the symmetries of the Hamiltonian (\ref{eq:hamiltonian}), the three tunneling amplitudes are equal. By defining $t_{0}\equiv\langle {w}_{\bm{0}A}|\hat{H}_{0}|w_{\bm{0}B}\rangle$, we can write
\begin{equation}
h^{(1)}_{12}(\bm{k})=t_{0}\left(1+e^{i\bm{k\!\cdot\! a_{1}}}+e^{-i\bm{k\!\cdot\! a_{2}}}\right)\equiv t_{0}Z_{0}(\bm{k})\equiv z(\bm{k})
\end{equation}
and $h^{(1)}_{21}(\bm{k})= z^{*}(\bm{k})$.  Finally, by defining 
\begin{equation}
t_{1\nu}=\langle {w}_{\bm{0}\nu}|\hat{H}_{0}|w_{\bm{(a_{1}+a_{2})}\nu}\rangle\equiv|t_{1\nu}|e^{i\varphi_{\nu}},
\end{equation}
and taking again into account the symmetries of the system (see Fig. \ref{fig:honeycomb}), the last term - corresponding to next-to-nearest tunneling between homologous sites - can be cast in the following form
\begin{align}
\label{eq:h2}
h^{(2)}_{\nu\nu}(\bm{k})&=|t_{1\nu}|\Big\{2\cos\left[\bm{k}\!\cdot\!(\bm{a}_{1}+\bm{a}_{2})
+\varphi_{\nu}\right]\\
&\qquad+
2\sum_{i=1,2}\cos\left(\bm{k}\!\cdot\!\bm{a}_{i}-\varphi_{\nu}\right)\Big\}
\equiv |t_{1\nu}|F_{\nu}(\bm{k}).
\nonumber
\end{align}
Notice that in general the onsite energies and the tunneling coefficients depend on the amplitudes of both the scalar and vector potentials: $E_{\nu}=E_{\nu}(s,\alpha)$, $t_{0}=t_{0}(s,\alpha)$, $|t_{1}|=|t_{1}|(s,\alpha)$, $\varphi_{\nu}=\varphi_{\nu}(s,\alpha)$. This is a direct consequence of the fact that the optimal choice for the basis of localized functions  $w_{\bm{j}\nu}(\bm{r})$ depends on the properties of overall structure of the Hamiltonian (\ref{eq:hamiltonian}).
By defining
\begin{equation}
\epsilon_{\nu}(\bm{k})=E_{\nu}+|t_{1\nu}| F_{\nu}(\bm{k}),
\end{equation}
we can write
\begin{equation}
h_{\nu\nu'}(\bm{k})=\left(\begin{array}{cc}
 \epsilon_{A}(\bm{k}) & z(\bm{k}) \\
 z^{*}(\bm{k}) & \epsilon_{B}(\bm{k})
\end{array}\right),
\label{eq:hmatrix}
\end{equation}
that is equivalent to the expression discussed in Ref. \cite{shao2008}.
However, we remark that here we have not made explicit use of the Peierls substitution, and that the dependence of Eq. (\ref{eq:h2}) on the phase $\varphi$ is a consequence of the symmetries of the full potential. 

Finally, by diagonalizing the matrix $h_{\nu\nu'}(\bm{k})$ and defining 
$f_{\pm}(\bm{k})\equiv(|t_{1A}|F_{A}(\bm{k})\pm |t_{1B}|F_{B}(\bm{k}))/{2}$, we get the following expression for the spectrum of the lowest two bands
\begin{equation}
\epsilon_{\pm}(\bm{k})=f_{+}(\bm{k})\pm\sqrt{\displaystyle{|\epsilon+f_{-}(\bm{k})|^{2}+|z(\bm{k})|^{2}}},
\end{equation}
that is a function of $|t_{0}|$, $|t_{1\nu}|$, and $\varphi_{\nu}$.

In the following we will consider for simplicity the degenerate case $\epsilon=0$ ($E_{A}=E_{B}$), corresponding to the potential in Eq. (\ref{eq:pot}). In this case, thanks to the symmetries of the system,  we have $|t_{1A}|=|t_{1B}|\equiv|t_{1}|$, $\varphi_{A}=-\varphi_{B}\equiv\varphi$ and $\langle {w}_{\bm{0}A}|\hat{H}_{0}|w_{\bm{0}B}\rangle=\langle {w}_{\bm{0}B}|\hat{H}_{0}|w_{\bm{0}A}\rangle$ (when $A$ and $B$ are equivalent the system is  invariant 
under rotation by $\pi$ radians around the center of any cell, see Fig. \ref{fig:honeycomb}). The latter implies that $t_{0}$ is real.
Remarkably, in this case the two tunneling amplitudes $t_{0}$ and $|t_{1}|$ and the phase $\varphi$ can be expressed in terms of specific properties of the spectrum. Let us start by noticing that $f_{+}(\bm{0})=6|t_1|\cos\varphi$, $f_{-}(\bm{0})=0$, $|z(\bm{0})|=3t_0$. In addition, we indicate with $\bm{k}_{D}$ the position of the Dirac points \cite{ibanez-azpiroz2013}, and define $\Delta_{\pm}\equiv\pm(\epsilon_{\pm}(\bm{0})-\epsilon_{\pm}(\bm{k}_{D}))$, that correspond to the two bandwidths when the tunneling coefficients satisfy the hierarchy $t_{1}\ll t_{0}$. Then, we have 
\begin{equation}
t_{0}=(\Delta_{+}+\Delta_{-}+\delta_{D})/{6},
\label{eq:t0}
\end{equation}
with $\delta_{D}\equiv\epsilon_{+}(\bm{k}_{D})-\epsilon_{-}(\bm{k}_{D})$ being the gap at the Dirac points, due to the presence of the vector potential.
Also, at e.g. $\bm{k}_D=(1,0)k_{L}$, we have
$f_{+}(\bm{k}_{D})=-3|t_1|\cos\varphi$, $f_{-}(\bm{k}_{D})=3\sqrt{3}|t_1|\sin\varphi$, $|z(\bm{k}_{D})|=0$, yielding \cite{shao2008}
\begin{equation}
\delta_{D}=6\sqrt{3}|t_1|\sin\varphi. 
\label{eq:gapbands}
\end{equation}
Another relation containing $|t_1|$ and $\varphi$ is 
\begin{equation}
\Delta_{+}-\Delta_{-}=18|t_{1}|\cos\varphi.
\label{eq:gapbands2}
\end{equation}
Then, by combining Eqs. (\ref{eq:gapbands}) and (\ref{eq:gapbands2}), we get
\begin{align}
\label{eq:t1}
|t_1|&=\frac{1}{18}\sqrt{(\Delta_{+}-\Delta_{-})^2+3\delta_{D}^2},\\
\varphi&=\tan^{-1}\left[\sqrt{3}\frac{\delta_{D}}{\Delta_{+}-\Delta_{-}}
\right].
\label{eq:phi}
\end{align}

Eqs. (\ref{eq:t0}), (\ref{eq:t1}) and (\ref{eq:phi}) represent an important contribution of this work: they provide a way to connect the value of the tunneling \textit{amplitudes} to gauge-invariant, measurable properties of the spectrum. Moreover, they also provide a straightforward method for computing the tunneling amplitudes, as the exact Bloch spectrum can be be readily computed by means of a standard Fourier decomposition \cite{lee2009,ibanez-azpiroz2013,*ibanez-azpiroz2013a}, even in the presence of a vector potential \cite{inpreparation}. 

\begin{figure}
\centerline{\includegraphics[width=1.05\columnwidth]{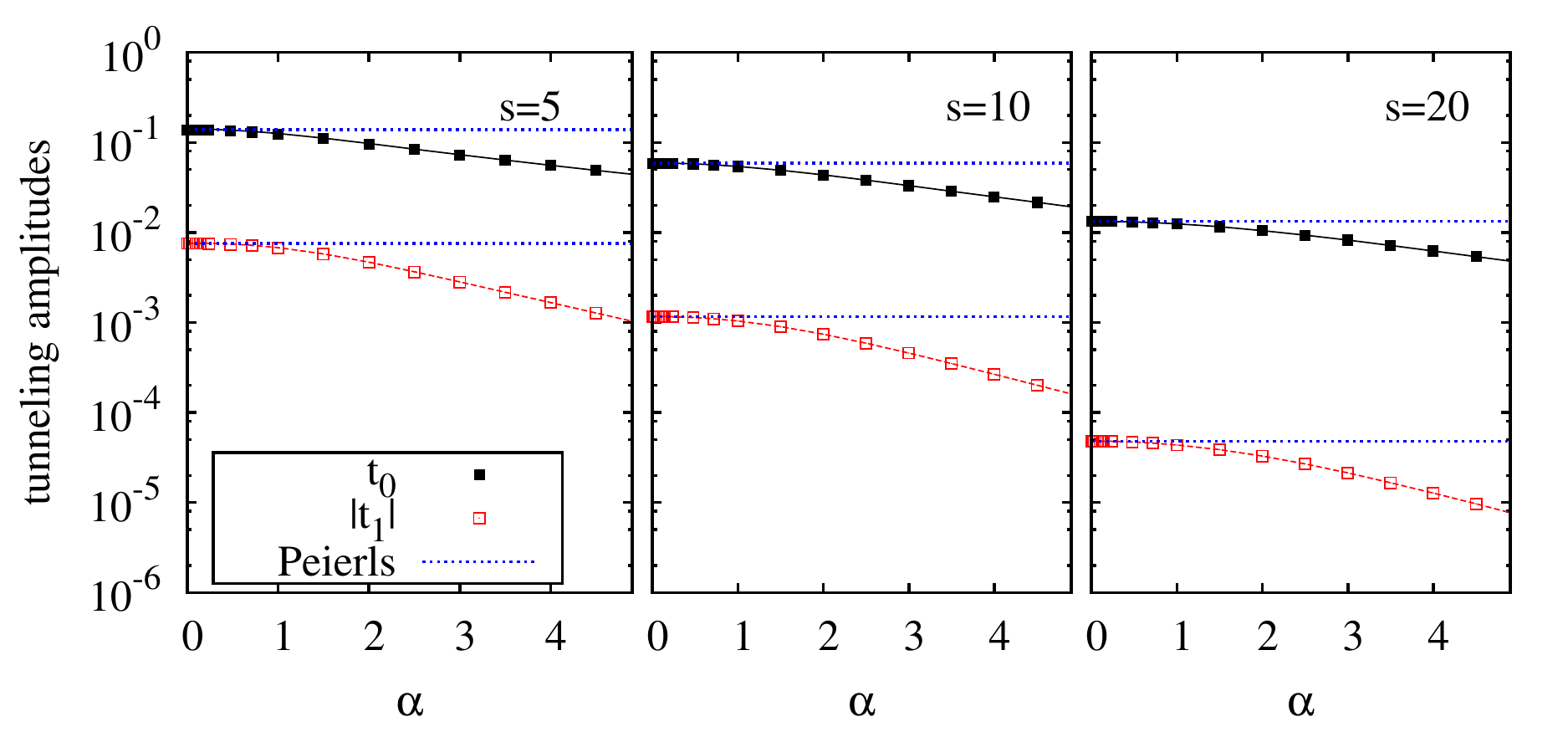}}
\caption{(Color online) Tunneling amplitudes $t_{0}$ (black full squares) and $|t_{1}|$ (red empty squares) for $s=5,10,20$ (from left to right),
as calculated from the MLWFs (points) and from the exact spectrum (lines). The agreement is remarkable. The horizontal blue (dotted) lines represent the values corresponding to the Peierls substitution. The values of the tunnelings are given in units of $E_{R}$.}
\label{fig:tun}
\centerline{\includegraphics[width=0.85\columnwidth]{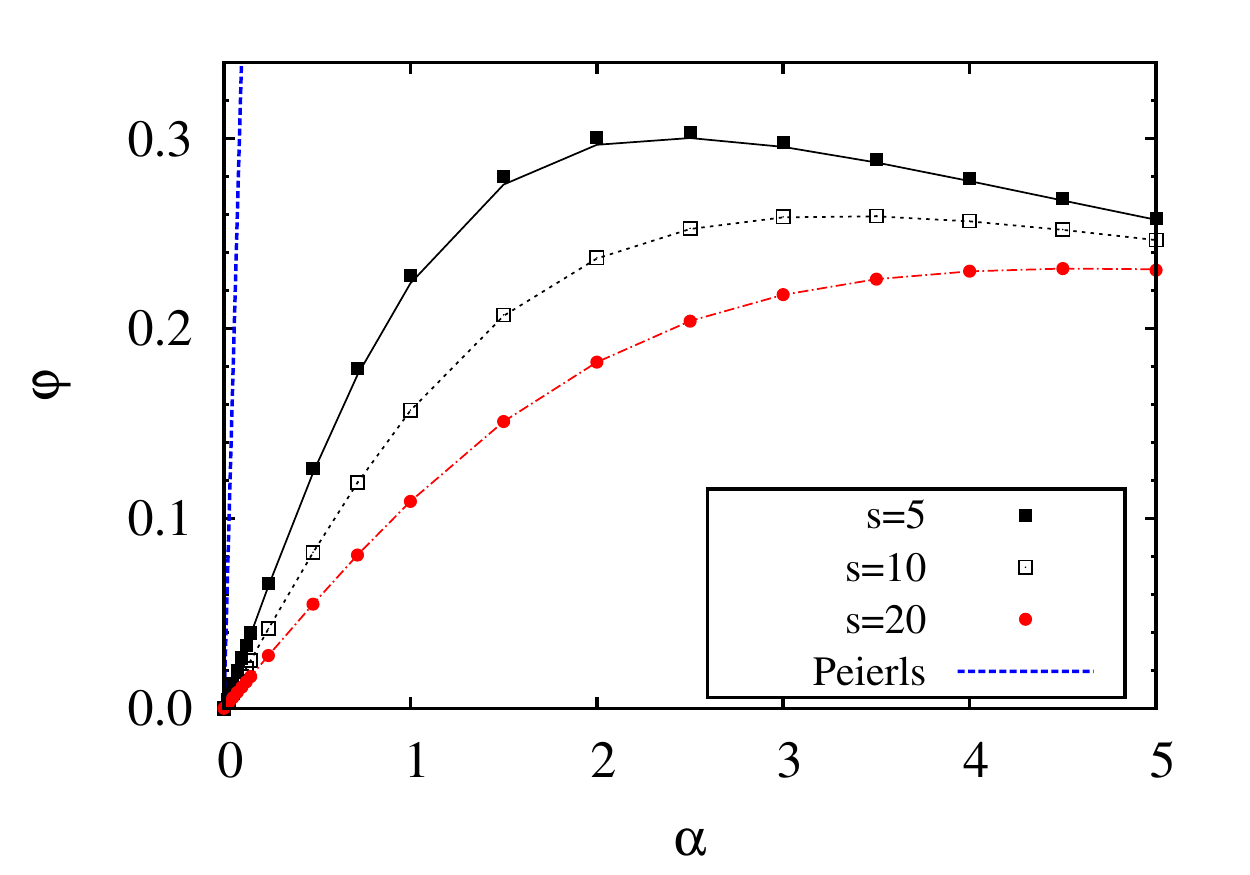}}
\caption{(Color online) Plot of the phase $\varphi$ as a function of the amplitude $\alpha$ of the vector potential, for $s=5,10,20$,
as calculated from the MLWFs (points) and from the exact spectrum (lines).
The prediction of the Peierls substitution is represented by the blue (dotted) line, that is almost indistinguishable from the vertical axis.}
\label{fig:phase}
\end{figure}

In addition, we compare these values with those computed \textit{ab-initio} from their definition in terms of the matrix elements $\langle w_{\bm{j}\nu}|{\hat{H}}_0|w_{\bm{j'}\nu'}\rangle$. To this end, we make use of the MLWFs for composite bands \cite{marzari1997,marzari2012}, which are defined through the following unitary mixing of the two lowest Bloch bands
\begin{equation}
w_{\bm{j}\nu}(\bm{r})=\frac{1}{\sqrt{S_{\cal B}}}
\int_{S_{\cal B}} \!\!\!\!d\bm{k} ~e^{-i\bm{k}\bm{R}_{\bm{j}}}\sum_{m=1}^{2}U_{\nu m}(\bm{k})\psi_{m\bm{k}}(\bm{r}),
\label{eq:mlwfs}
\end{equation}
with $\bm{R}_{\bm{j}}\in{\cal{B}}$, $\psi_{m\bm{k}}$ being the eigenfunctions of the Hamiltonian (\ref{eq:hamiltonian}) \cite{kohn1959}, and  $U_{\nu m}(\bm{k})$ a 
$2\times 2$ unitary matrix, periodic in $\bm{k}$-space,
which minimizes the spread of $w_{\bm{j}\nu}(\bm{r})$ \cite{marzari1997}. In the present case, the MLWFs are obtained by modifying the code discussed in Ref. \cite{ibanez-azpiroz2013,*ibanez-azpiroz2013a} in order to include a vector potential. 
The MLWFs turn out to be complex due to the breaking of time-reversal, and this explains the emergence of a phase factor in the tunneling coefficients \cite{inpreparation}.
The values obtained for $t_{0}$, $|t_{1}|$ and $\varphi$ are shown in Figs. \ref{fig:tun}, \ref{fig:phase}, along with those extracted from the spectrum. The agreement is remarkable \footnote{We have verified that values of $t_{0}$, $|t_{1}|$ and $\varphi$ obtained with the two methods allow to reproduce the exact spectrum with great accuracy in the proper tight-binding regime, $s\gtrsim3$; a detailed discussion will be presented elsewhere \cite{inpreparation}.}.

\begin{figure}
\centerline{\includegraphics[width=0.8\columnwidth]{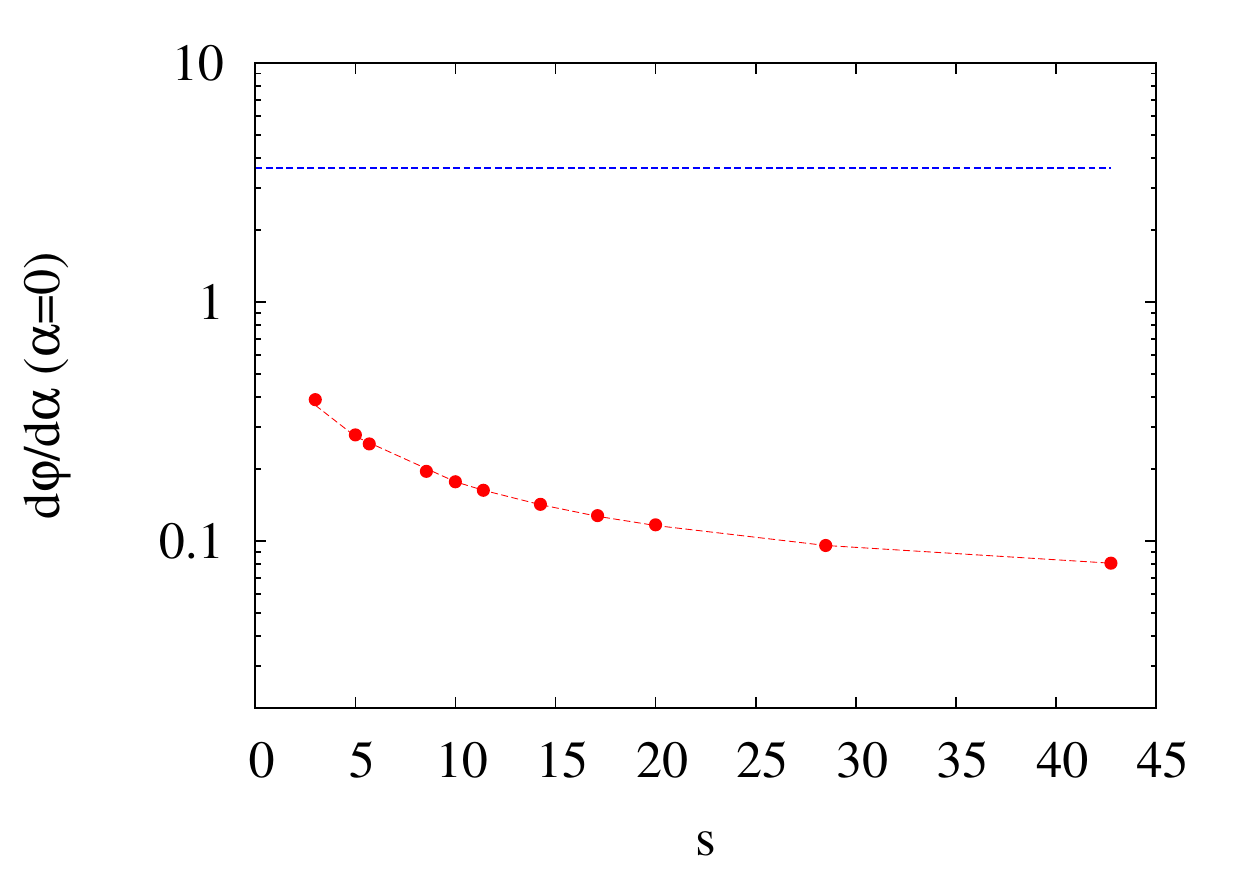}}
\caption{(Color online) Plot of $d\varphi/d\alpha|_{\alpha=0}$ calculated from the MLWFs (points) and from the exact spectrum (lines), as a function of the amplitude $s$ of the honeycomb potential. The horizontal dashed line represents the value corresponding to the Peierls phase $\varphi=({2\pi}/{\sqrt{3}})\alpha$. Note the logarithmic scale on the vertical axis. We remark that the present tight-binding model with up to nearest-neighbor tunnelings is accurate only for  $s\gtrsim3$; for lower values it may be necessary to consider also other next-to-leading tunneling coefficients \cite{ibanez-azpiroz2013,*ibanez-azpiroz2013a}.}
\label{fig:dphase}
\end{figure}

From these figures we can identify two regimes as a function of the amplitude $\alpha$ of the vector potential: \textit{(i)} for small enough values, $\alpha\lesssim1$, where $t_{0}$ and $|t_{1}|$ are almost constant and the phase $\varphi$ is linear in $\alpha$; \textit{(ii)} for $\alpha\gtrsim1$ where the dependence on $\alpha$ is less trivial. In particular, in the latter regime, $t_{0}$ and $|t_{1}|$ present a pronounced dependence on $\alpha$, in clear contrast with the Peierls substitution (see horizontal lines in Fig. \ref{fig:tun}) which assumes the phase $\varphi$ to be the only $\alpha$-dependent quantity. However, this dependence is not surprising, as the presence of the vector potential may significantly affect both the Bloch eigenfunctions $\psi_{m\bm{k}}$ \cite{kohn1959} and the gauge transformation $U_{\nu m}$ entering Eq. (\ref{eq:mlwfs}) \cite{alexandrov1991a}, so that the usual implicit assumption that the basis of localized orbitals is not affected by the vector potential (see e.g. \cite{boykin2001}) is generally not valid. On the other hand, the calculated phase strongly deviates from the linear behavior expected from the Peierls substitution, namely $\varphi=\int_{\bm{r}_{A}}^{\bm{r}_{A}-\bm{a}_{1}}\bm{A}\cdot d\bm{r}=({2\pi}/{\sqrt{3}})\alpha$ \cite{shao2008}, see Fig. \ref{fig:phase}. This figure reveals that the Peierls substitution dramatically fails even in the ``linear'' regime, as it predicts a slope for the phase far much larger than the actual one. Moreover, it completely neglects its dependence on the amplitude $s$ of the scalar potential (that is appreciable even in the full tight-binding regime, $s>10$). This is particularly evident from Fig. \ref{fig:dphase}, where we plot the behavior of the angular coefficient in the linear regime, $d\varphi/d\alpha|_{\alpha=0}$, as a function of $s$. This figure provides further evidence that the Peierls substitution does not even provide a reasonable estimate for the order of magnitude of $\varphi$ in the linear regime.
Essentially, the reason for the breakdown of the Peierls substitution resides in the fact that the hypotheses under which it has been rigorously demonstrated \cite{luttinger1951,boykin2001} cannot be satisfied in the Haldane model. Most importantly, the vector potential can not be considered as slowly varying \cite{graf1995}, as it varies on the same length scale as the lattice (see Fig. \ref{fig:honeycomb}). As a consequence, both the scalar and vector potentials must be treated on equal foot, and all parameters ($t_{0}$, $|t_{1}|$ and $\varphi$) must be considered as dependent on both $s$ and $\alpha$. 

In summary, we have presented two independent calculations of the tight-binding parameters for the Haldane model with ultracold atoms \cite{shao2008}, one based on their \textit{ab-initio} definition in terms of the MLWFs, and the other in terms of gauge invariant properties of the spectrum, summarized in Eqs. (\ref{eq:t0}), (\ref{eq:t1}) and (\ref{eq:phi}). The latter provides a straightforward approach whenever the spectrum can be measured or computed with sufficient accuracy.
The results obtained with the two methods present a remarkable agreement, and demonstrate the inadequacy of the Peierls substitution, which fails in predicting quantitative and even qualitative properties of the system.
The reason for this breakdown is due to the fact that the regime of validity of the Peierls substitution cannot be fulfilled in any realization of Haldane model, regardless of the system, being it cold atoms in optical lattices or electrons in a solid. Our results indicate that a careful revision of the validity of the commonly employed Peierls substitution in tight-binding models is necessary.

\textit{Acknowledgments.}
This work has been supported by the UPV/EHU under programs 
UFI 11/55 and IT-366-07, the Spanish Ministry of
Science and Innovation through Grants No. FIS2010-19609-C02-00 and FIS2012-36673-C03-03, and the Basque Government through the Grant No. IT-472-10.
JIA would like to acknowledge support from the HGF-YIG Programme VH-NG-717 (Functional Nanoscale Structure and Probe Simulation Laboratory-Funsilab).


%

\end{document}